\documentclass[utf8]{FrontiersinHarvard}

\usepackage{url,hyperref,lineno,microtype}
\usepackage[onehalfspacing]{setspace}
\usepackage{graphicx}
\usepackage{booktabs}
\usepackage{array}
\usepackage{amsmath}
\usepackage{textcomp}
\usepackage{xcolor}

\graphicspath{{figures/}}

\def\keyFont{\fontsize{8}{11}\helveticabold }

\def\firstAuthorLast{Ismail {et~al.}}
\def\Authors{Nuzla Ismail\,$^{1,*}$, Shanika Amarasoma\,$^{2}$, Michelle Galeas-Pena\,$^{1}$ and Sarah C. Glover\,$^{1}$}
\def\Address{$^{1}$Department of Medicine, Section of Gastroenterology, Tulane University, New Orleans, LA, United States \\$^{2}$Independent Researcher, Advanced Analytics and AI, Colombo 00900, Sri Lanka}
\def\corrAuthor{Nuzla Ismail}
\def\corrEmail{fismail@tulane.edu}

\begin{document}

\onecolumn
\firstpage{1}

\title[Truecell: high-fidelity Seurat analysis in Python]{Truecell
		reproduces R Seurat's single-cell analysis outputs natively in the Python
		ecosystem}

\author[\firstAuthorLast ]{\Authors}
\address{}
\correspondance{}
\extraAuth{}

\maketitle

\begin{abstract}
	\section{}
Single-cell RNA-sequencing analyses predominantly use either Seurat (R) or Scanpy (Python), with the choice often driven by programming language preference. However, by default, these tools produce divergent variable features, neighbour graphs, clusters, and marker genes. Laboratories requiring Python, which supports most deep-learning, foundation-model, and agent tools, must re-implement Seurat analyses in a framework that does not replicate Seurat's results. Truecell was developed as a Python implementation of the Seurat interface, preserving function names, arguments, default settings, and the object model to facilitate seamless transfer of analyses. In eighteen paired end-to-end evaluations against R Seurat, deterministic outputs matched to floating-point precision, and fold-change order was identical across all nine differential expression tests. In a three-arm benchmark on three datasets, with fixed user parameters and 20 seeds per tool, Truecell more closely reproduced Seurat's clustering than a Seurat-configured Scanpy in all 12 combinations of dataset and resolution settings. Truecell's marker genes and enriched pathways were also closer to Seurat's than Scanpy's were, and pseudobulk DESeq2 reproduced Seurat's gene lists with a Jaccard index ranging from 0.95 to 1.00. This agreement reflects fidelity to Seurat rather than biological correctness.
	\tiny
	\keyFont{\section{Keywords:} single-cell RNA sequencing, Seurat, Scanpy,
		Python, reproducibility, benchmarking, differential expression,
		spatial transcriptomics}
\end{abstract}

\section{Introduction}
Single-cell RNA sequencing has become a standard experimental method.
Two software toolkits dominate the computational analysis that follows.
Seurat \citep{satija2015,butler2018,stuart2019,hao2021,hao2024}, written in R,
and Scanpy \citep{wolf2018}, written in Python, together account for the
majority of published single-cell analyses \citep{rich2026}. Both implement the
same general workflow: filter cells and genes, normalize data, select variable
features, reduce dimensions, build a neighbour graph, cluster, and test for
differentially expressed markers. The selection between these toolkits is
typically presented to students and collaborators as a decision based primarily
on programming language preference. This perspective assumes that both toolkits
produce equivalent results, a premise that requires further examination.

\citet{rich2026} examined this assumption. Running both packages on the same count matrices
at their own defaults, they found the selected variable features overlapped
at a Jaccard index of 0.22, the shared-nearest-neighbour graphs at a median
of 0.11, the clusterings at an adjusted Rand index of 0.53, and the significant
marker genes at 0.62, with Seurat reporting roughly half again as many
markers. The magnitude of this difference was calibrated experimentally:
it is comparable to the variability introduced by sequencing fewer than 5\% of
reads or analysing fewer than 20\% of the cells. Some of the
difference closes when arguments are matched, but some remains.
Rewriting an established Seurat analysis in Scanpy is therefore not a
translation of the analysis but a re-derivation of it, and the results are not
guaranteed to remain consistent after this change.

Nonetheless, there are strong reasons for laboratories to conduct analyses
using Python. Most deep-learning software used in single-cell biology is written in Python, on more than one framework: scVI and scvi-tools are built
on PyTorch \citep{paszke2019,lopez2018,gayoso2022}, the deep count autoencoder
DCA on Keras and TensorFlow \citep{abadi2016,eraslan2019}, and the current
single-cell foundation models, among them scGPT, Geneformer, scFoundation and
scBERT, on PyTorch \citep{cui2024,gulati2025}. Agent frameworks and large language model client libraries are increasingly integrated into analysis workflows and are typically implemented in Python. Research groups whose analytical processes are based in R, but whose subsequent steps require Python,
must either maintain two separate pipelines, rewrite the existing pipeline, or accept that transitioning the pipeline to Python may alter its outputs.

\begin{figure}[tbp]
	\begin{center}
		\includegraphics[width=\linewidth,height=0.8\textheight,keepaspectratio]{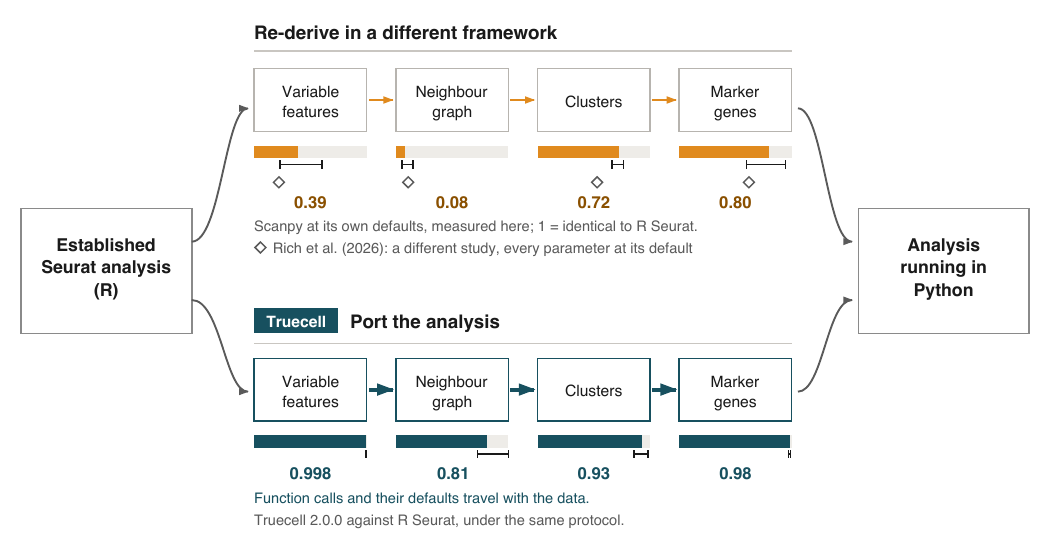}
	\end{center}
	\caption{\textbf{Two ways to move a Seurat analysis into Python,
		measured under one protocol.} Each bar is agreement with R Seurat 5.5.1
		at one stage of the standard workflow, where 1 means identical output: the
		median over PBMC 3k, ifnb and panc8, with the line beneath it spanning the
		three. Upper track: the analysis re-derived in Scanpy 1.12.3 at its own
		defaults. Lower track: the same calls in Truecell 2.0.0. User-set parameters
		are the same in both, and the metrics are those of
		Figure~\ref{fig:benchmark}. Open diamonds are the values \citet{rich2026}
		reported for Seurat against Scanpy with every parameter at its default, a
		different protocol.}
	\label{fig:two-paths}
\end{figure}

Here we present Truecell, a Python port of Seurat's
analysis interface designed to eliminate this trade-off. Truecell is a port rather
than a reimplementation: it preserves Seurat's function names, argument
names, default values and object model, so a Seurat user can read and write
Truecell code without learning a new framework, and existing analyses can be
transferred with minimal rewriting. Figure~\ref{fig:two-paths} contrasts the
two routes into Python, both measured in this study: a re-derivation,
which alters the results at each stage where the frameworks make different
choices, versus a port, which transfers the function calls and their defaults,
along with the data. We validate Truecell against R Seurat across eighteen
paired end-to-end evaluations, benchmark it against Scanpy on three datasets
under a protocol matched to \citet{rich2026}, test whether the remaining
differences change pathway enrichment and pseudobulk differential expression,
verify spatial data handling on a public Xenium section, and measure what
reproducing Seurat costs in runtime and memory relative to Seurat itself.

Every agreement reported here is fidelity to R Seurat as a reference
implementation, not a claim of biological correctness (Section~\ref{sec:limitations}).

\section{Materials and Methods}

\subsection{Implementation}
\label{sec:m-implementation}

Truecell is implemented in Python and relies on NumPy \citep{harris2020} and
SciPy \citep{virtanen2020} for numerical computation, and on scikit-learn for
dimensionality reduction and exact nearest-neighbour search, together
with the standard Python libraries for data visualization and data
management. Continuous integration runs the tests on Linux under
Python 3.12 and 3.13, the tutorials against the built wheel, and, weekly, the
quick-start on macOS arm64. Sparse
matrices are used throughout. An optional on-disk layer enables support
for matrices that exceed the available memory.

\subsection{Validation strategy}
\label{sec:m-validation}

Each of the eighteen evaluations pairs a Python script with a corresponding R script
that executes the same analysis (Table~\ref{tab:evals}). Both scripts perform the
Seurat analysis on identical input data. The comparison is performed using an
\emph{anchor}, defined as a named quantity computed by both scripts and
evaluated with a specified tolerance. Examples of anchors include an individual
cell's cluster assignment, a gene's fold change or fitted model parameters. This
approach is preferred over summary statistics, which could agree
coincidentally. Cells and genes are matched by barcode and symbol, rather than
by position. Each randomized step is seeded in both languages, and each
evaluation states its tolerance and the source of that tolerance.

Every value comes from Truecell 2.0.0 installed from PyPI, R 4.6.1 with
Seurat 5.5.1 and SeuratObject 5.4.0, and Scanpy 1.12.3 (Supplementary Table S2).
The three-arm benchmark and the enrichment, pseudobulk and spatial analyses followed a plan written before they were run, and analyses added after results had been seen are labelled exploratory
(Supplementary Note~S1).

Principal components, independent components and cluster labels are each
arbitrary up to sign, order or name, so they are matched between the tools
before they are scored. A principal component is fixed only up to sign, and
among near-degenerate eigenvalues only up to order, so the two sets of components
are paired one to one on absolute correlation by optimal assignment
\citep{kuhn1955}; the reported $|r|$ is the mean over that pairing, and whether
the optimal assignment was the identity permutation is recorded. On PBMC
3k it was, across all ten dimensions the clustering runs on. Independent
components are handled in the same manner. Cluster labels are arbitrary in both
tools; therefore, partitions are scored by the adjusted Rand index (ARI;
\citealt{hubert1985}), which is label-invariant; a Hungarian matching of the
contingency table is used only to report the per-cell concordance alongside it.

UMAP \citep{mcinnes2018} coordinates are not scored in any reported agreement metric.
\texttt{run\_umap} is validated as an implementation correspondence
instead: it uses the same algorithm, the same parameters and the
same input embedding, and it is timed, but its output is not compared. A
UMAP embedding is not identified up to anything stronger than its topology: the
optimization is stochastic and non-convex, R Seurat drives \texttt{uwot} where
Truecell drives \texttt{umap-learn}, and the two differ in initialization and in
optimization detail, so two runs of the \emph{same} library under different
seeds already disagree on the coordinates. A coordinate-level agreement number
would therefore be measuring the seed and the library, and a high one would be a
surprising result. The weight is carried on either side of the embedding
instead: the neighbour graph it is computed from is compared directly, and the
cluster assignments drawn from it are scored by the adjusted Rand index.

\subsection{Rank, threshold and parameter-range comparisons}
\label{sec:m-readable}

Four comparisons were added to the paired evaluations because a set overlap
and a worst-case bound answer neither of the two questions someone reads a
marker table to answer. All four are Truecell against R Seurat 5.5.1
alone; none of them is a three-arm comparison. Their results are
reported in Section~\ref{sec:readable} and Table~\ref{tab:readable}.

\emph{Differential expression.} The nine \texttt{find\_markers} tests
already run on one shared cluster assignment, so no clustering difference can
present itself as a differential expression difference. On the genes both arms
return, the added scoring is Spearman $\rho$ and Kendall $\tau$ of
\texttt{avg\_log2FC}; the top-50 overlap ranked by $|\log_2\mathrm{FC}|$ rather
than by $p$; agreement of \texttt{p\_val\_adj} to six significant figures; and
agreement at a significance level of 0.05. Two measurement artefacts
were removed first. Seurat clamps \texttt{p\_val\_adj} at 1, and for wilcox 11,948 of
13,714 genes land there, so an unconditioned fraction-identical reads about
0.87 before a single informative gene is considered, and the unclamped
subset is scored separately. Neither language round-trips a float64 through
a CSV at its defaults: R's \texttt{write.csv} renders 15 significant figures and
R's own \texttt{sprintf} is not correctly rounded at higher precision, while
pandas misparses roughly a third of random doubles by one unit in the last
place. The R script therefore also writes a C99 hex-float table, and the Python
side reads the table using round-trip parsing. Before that change, an ``are these
identical'' comparison partly measured the two languages' text formatters.

\emph{Resolution range.} Choosing a clustering resolution means running several
and comparing them, so both arms use the vector idiom,
\texttt{find\_clusters(pbmc, resolution=[0.4, 0.8, 1.2, 0.5])} and
\texttt{FindClusters(pbmc, resolution = c(...))}, and every resolution is
scored against R. Each is clustered from the same seed rather than from a
running stream, so a partition does not depend on what preceded it. The
value 0.5 is given last because Seurat leaves the object on the last resolution
in the sequence, and every step below the clustering call is written against
it.

\emph{Gene-program scoring.} \texttt{AddModuleScore} draws its control genes
at random, and R's generator is not NumPy's, so at the default settings
correlation is the ceiling, and exact agreement is unavailable. One regime
removes the draw: \texttt{nbin}~$=$~1 puts every gene in a single bin, and
\texttt{ctrl} equal to the pool size then draws that whole bin, so
\texttt{sample(n, n)} is a permutation, the control \emph{set} is forced and
only its summation order is free. Both regimes are reported, since the
second is what separates a faithful port from one whose binning is subtly
incorrect.

\emph{Group summaries.} Truecell's \texttt{aggregate\_expression} and
\texttt{average\_expression} are pinned against \texttt{AggregateExpression}
and \texttt{AverageExpression} on a hard-coded count matrix, so neither
language's generator is in the loop and both sides are provably reading the
same numbers.

\subsection{Three-arm benchmark}
\label{sec:m-benchmark}

R Seurat 5.5.1 (R 4.6.1) served as the reference; the comparison arms were
Scanpy 1.12.3 and Truecell 2.0.0. All arms read the same count matrices and
applied the same construction filter (genes in at least 3 cells, cells with at
least 200 genes, both computed on the raw matrix as \texttt{CreateSeuratObject}
does). For PBMC 3k, a standard tutorial quality filter was also applied 
($200 < \text{nFeature} < 2{,}500$, mitochondrial fraction $<$ 5\%).
Normalization was performed using log-normalization with a scale factor of $10^4$. The
parameters a user sets explicitly were identical across arms: 2,000 variable
features, 50 principal components, the first 10 dimensions, $k = 20$
nearest neighbours per cell (counting the cell itself, as Seurat's
\texttt{k.param} does), and a clustering resolution of 0.5 in the canonical
run. Differential expression used the Wilcoxon test, positive markers only,
with \texttt{min.pct} 0.1, a log-fold-change threshold of 0.25 and an adjusted
$p$-value cutoff of 0.05 applied uniformly at comparison time.

Four metrics were computed against the reference arm, chosen to match those
reported by \citet{rich2026}: the Jaccard index of the selected
variable-feature sets; the median over cells of the per-cell neighbour-set
Jaccard index, excluding self; the adjusted Rand index of the clusterings over
shared barcodes; and the Jaccard index of the union of per-cluster positive
markers.

Because the arms search for neighbours differently \citep{dong2011}, the
neighbour metric is scored against an exact search on Seurat's PCA embedding,
which Seurat's own graph matches at a median Jaccard index of 1.000
(Supplementary Note~S3).

Scanpy was run in two configurations. The \emph{defaults} arm used its own
dispersion-based variable-feature method and Leiden clustering
\citep{traag2019}. The \emph{matched} arm used its \texttt{seurat\_v3}
variable-feature method and Louvain clustering \citep{blondel2008} at the same
resolution, driven through igraph because Scanpy's own Louvain entry point
could not be resolved in the environment used in this study. The igraph function
\texttt{community\_multilevel} draws on igraph's own random-number generator
rather than NumPy's and is seeded explicitly via
\texttt{igraph.set\_random\_number\_generator}. Differential expression in both
Scanpy arms was computed on the log-normalized matrix rather than the
scaled one, matching what \texttt{FindAllMarkers} tests.

Clustering is randomized in every arm, Seurat included, so each arm
clustered with seeds 1 to 20 at resolutions 0.4, 0.5, 0.8 and 1.2, and a seed's
agreement is the mean adjusted Rand index over Seurat's 20 seeds. Seurat's
self-agreement, each of its seeds against the other 19, is the ceiling.

In each of the 12 dataset and resolution settings, Truecell's seeds were
compared with those of each Scanpy arm and with Seurat's self-agreement by a two-sided
Mann--Whitney U test \citep{mann1947} with the Hodges--Lehmann shift
\citep{hodges1963}, Holm-adjusted \citep{holm1979} across the 12 settings
(Supplementary Notes~S2 and~S3).

To separate the marker test from clustering, each arm's test was
also run on Seurat's most representative partition at resolution 0.5, and on each
arm's own clustering from seeds 1 to 10 (Supplementary Note~S3).

\subsection{Pathway enrichment and pseudobulk differential expression}

\label{sec:m-downstream}

Downstream readouts were compared by applying one implementation to every
arm's output. Each cluster's
markers were tested for over-representation with GSEApy 1.3.1 \citep{fang2023}
against the MSigDB Hallmark sets \citep{subramanian2005,liberzon2015}, Gene
Ontology biological processes \citep{ashburner2000,go2023} and the C8 cell-type
signatures, and clusters matched to Seurat's were scored by the Jaccard index of
significant terms. On ifnb, pseudobulk differential expression followed Seurat's
vignette: counts were summed by condition, donor and cell type \citep{kang2018},
and stimulated and control samples were compared with the DESeq2 test
\citep{love2014}, through PyDESeq2 in Truecell \citep{muzellec2023}
(Supplementary Note~S3).

\subsection{Spatial transcriptomics}
\label{sec:m-spatial}

The Xenium path was also tested on the public 10x Genomics human breast
cancer section \citep{janesick2023}, comparing cells, genes, counts and centroids
exactly, principal components as above and clusterings from five seeds, and
interoperability was tested through AnnData \citep{virshup2023} and SpatialData
\citep{marconato2025} (Supplementary Note~S3).

\subsection{Performance benchmark}
\label{sec:m-performance}

Runtime and memory were measured by a separate suite in which the Truecell and
R Seurat scripts are line-for-line counterparts, running the same pipeline on
the same inputs with the same parameters. It times the
standard workflow on PBMC 3k, PBMC 8k, ifnb and THP-1 (2,700 to 20,729
cells) and, separately, SCTransform \citep{hafemeister2019} and the differential expression tests on PBMC 3k, integration on ifnb and spatial statistics on the Xenium mouse brain section.
It ran on an Apple M5 Pro (18 cores, 48~GB, macOS 27.0) and on a virtual
Linux x86-64 server (8 AMD EPYC cores, 23~GB, Ubuntu 24.04), with both language
runtimes linked to an optimized BLAS on each (Supplementary Note~S3).

Time is measured inside each process and memory from outside it: R's garbage
collector and Python's allocator account for memory differently and neither
observes the other, so a figure reported by either runtime would not be
comparable. Instead, the parent process samples the resident-set size of the
child's process tree every 50~ms and intersects those samples with the step
boundaries the child logs, so both arms are measured by one instrument on one
clock. Reported per-step figures are peaks during that step; the pipeline figure
is the peak for the whole process, which is what decides whether a machine
can run the workload. Each benchmark was run four times, or three times for
the heavier operations. The first run warms up compiled kernels and lazily
loaded packages and is discarded, and the median of the remaining runs is
reported.

Every step records an anchor (Section~\ref{sec:m-validation}), such
as cells kept, clusters found, genes tested or markers returned, so that
a timing difference cannot be a work difference in disguise. These counts
agree at every step on both machines.

Truecell and Seurat do not always find the same number of clusters, and
one-versus-rest marker detection runs one test per cluster, so a difference in
cluster number would change how much work marker detection does. The Truecell
arm therefore runs first and writes its cell-to-cluster assignment, and the R
arm reads that assignment before marker detection. Interpreter start-up and R's
\texttt{library()} calls are recorded but excluded from pipeline totals, as are
steps that only one arm can perform; these are reported separately. The
remaining total, the steps both arms share, is reported as the
\emph{shared pipeline}.

\subsection{Datasets}
\label{sec:datasets}

The analyses use PBMC 3k and PBMC 8k (10x Genomics), the interferon-$\beta$ stimulation
dataset \citep[ifnb;][]{kang2018}, the panc8 compendium of eight human islet
datasets from five technologies
\citep{baron2016,muraro2016,segerstolpe2016,lawlor2017,grun2016}, CBMC CITE-seq
\citep{stoeckius2017}, cell hashing \citep{stoeckius2018}, THP-1 ECCITE-seq
\citep{papalexi2021}, 10x Xenium and Visium mouse brain sections, and the 10x
Xenium human breast cancer section \citep{janesick2023}. All are public. The package downloads and caches the 10x PBMC, CITE-seq, cell-hashing, ECCITE-seq and mouse brain datasets on first use; ifnb and panc8 are exported once from SeuratData by \texttt{tutorials/export\_seuratdata.R} in the repository, and the breast cancer section is read from the 10x Genomics download. Supplementary
Table~S13 gives their sizes and the analyses that use each; the three-arm
benchmark uses PBMC 3k, ifnb and panc8.

\subsection{Use of generative AI}
\label{sec:m-genai}

Generative AI was used to write code and draft text, as the Generative AI
statement in the back matter describes. Every value reported was computed by the
paper's analysis scripts.

\section{Results}
\label{sec:results}

\subsection{Truecell reproduces Seurat's analysis surface in Python}
\label{sec:surface}

Truecell exposes 90 public functions covering the Seurat v5 workflow, of
which 17 are plotting functions and five set plot themes and palettes,
plus 22 classes that implement the object model. The correspondence with R is
deliberately mechanical for function and argument names: \texttt{FindMarkers}
becomes \texttt{find\_markers}, \texttt{RunPCA} becomes \texttt{run\_pca},
\texttt{NormalizeData} becomes \texttt{normalize\_data}, and argument names
follow the same substitution
(\texttt{logfc.threshold}~$\rightarrow$~\texttt{logfc\_threshold}). Default
values are carried across unchanged, and a test now fails on any default
that differs from the mirrored Seurat function's without a recorded reason. It
found nine differences, among them Seurat~4's marker thresholds in
\texttt{find\_markers}, all corrected in 2.0.0. Defaults matter more than the naming: the differences
\citet{rich2026} measured between Seurat and Scanpy arise substantially from
defaults that were each chosen sensibly and independently.

The object model is ported as well as the functions, though its accessor names
follow Python's naming convention rather than a literal transliteration of R's:
a Truecell object holds named assays, layered expression matrices in the v5
arrangement, dimensional reductions, neighbour graphs, per-cell metadata and a
command log, and supports the operations a Seurat user expects:
\texttt{cell\_names} (R: \texttt{Cells}), \texttt{feature\_names}
(\texttt{Features}), \texttt{embeddings} (\texttt{Embeddings}),
\texttt{fetch\_data} (\texttt{FetchData}), \texttt{idents} (\texttt{Idents}),
\texttt{which\_cells} (\texttt{WhichCells}) and subsetting. A dedicated
evaluation of the container alone, independent of any algorithm, matched R on
91 of 91 checked properties (Table~\ref{tab:evals}).

\begin{table}[tbp]
	\footnotesize
	\caption{\textbf{The eighteen paired evaluations, grouped by analysis
		domain.} Each runs the same analysis in R Seurat 5.5.1 and in Truecell
		2.0.0 and compares the outputs at the finest granularity the analysis
		allows. The headline results are shown; Supplementary Data Sheet 1 lists every
		anchor with its tolerance. ARI, adjusted Rand index; PC, principal
		component.}
	\label{tab:evals}
	\begin{center}
		\begin{tabular}{@{}>{\raggedright\arraybackslash}p{0.17\linewidth}>{\raggedright\arraybackslash}p{0.27\linewidth}>{\raggedright\arraybackslash}p{0.50\linewidth}@{}}
			\toprule
			\textbf{Domain} & \textbf{Evaluation (dataset)} & \textbf{Agreement with R Seurat} \\
			\midrule
			Objects and preprocessing & Object model (PBMC 3k) & 91 of 91 anchors \\
			 & SCTransform (PBMC 3k) & model parameters Spearman 1.0000; the same 3,848 Poisson genes \\
			 & Out-of-core matrices (PBMC 3k) & 14 of 14 anchors; on-disk and in-memory results bit-identical \\
			\addlinespace
			Dimensions and clusters & Guided clustering (PBMC 3k) & same 2,638 cells; 1,998 of 2,000 variable features; PCA $|r|$ 0.9986; ARI 0.928 (9 and 9 clusters) \\
			 & Dimensional reduction (PBMC 3k) & ICA $|r|$ 0.9991; 14 and 13 PCs kept, within the declared band of $\pm2$ \\
			 & Subclustering (PBMC 8k) & global ARI 0.983; T and NK subclusters ARI 0.934 \\
			 & Leverage sketching (ifnb) & leverage scores Spearman 1.000000 in the exact regime \\
			\addlinespace
			Differential expression and scores & Differential expression (PBMC 3k) & fold changes to $1.8\times10^{-15}$; fold-change order identical in all nine tests (Table~\ref{tab:readable}) \\
			 & Cell cycle and module scores (THP-1) & 95.9\% of phase calls; scores Pearson $\geq$ 0.997; $6.7\times10^{-15}$ with the control set forced \\
			\addlinespace
			Multimodal and perturbation & CITE-seq WNN (CBMC) & protein weights Pearson 0.984; 98.7\% of cell-type labels \\
			 & Cell hashing (PBMC) & 99.8\% of HTODemux calls; 94.7\% of MULTIseqDemux calls \\
			 & Mixscape (THP-1) & 97.7\% of perturbation classes \\
			\addlinespace
			Integration and mapping & Batch integration (ifnb) & batch mixing 0.991, 0.992 and 0.917 after Harmony, CCA and RPCA (R: 0.991, 0.991 and 0.917) \\
			 & Anchor internals (ifnb) & 99.9\% of Seurat's CCA anchors and 100\% of its RPCA anchors \\
			 & Reference mapping (panc8) & 98.9\% of predicted labels \\
			\addlinespace
			Spatial & Xenium spatial (mouse brain) & every deterministic anchor \\
			 & Spatial statistics (mouse brain) & 39 of 39 anchors; Moran's~I to $1.6\times10^{-14}$ \\
			 & Visium (mouse brain) & 24 of 24 anchors; spot coordinates identical \\
			\bottomrule
		\end{tabular}
	\end{center}
\end{table}

\subsection{Eighteen paired evaluations quantify agreement with R Seurat}
\label{sec:evals}

Each evaluation was performed twice: once using R Seurat and once using
Truecell. The outputs were then compared at the finest granularity the analysis
permits---per cell, per gene or per fitted parameter---rather than by
summary statistics. Table~\ref{tab:evals} lists the eighteen evaluations with
datasets and headline agreements, Supplementary Data Sheet 1 gives every
anchor, and Supplementary Table S1b compares the two machines.

\begin{figure}[tbp]
	\begin{center}
		\includegraphics[width=\linewidth,height=0.8\textheight,keepaspectratio]{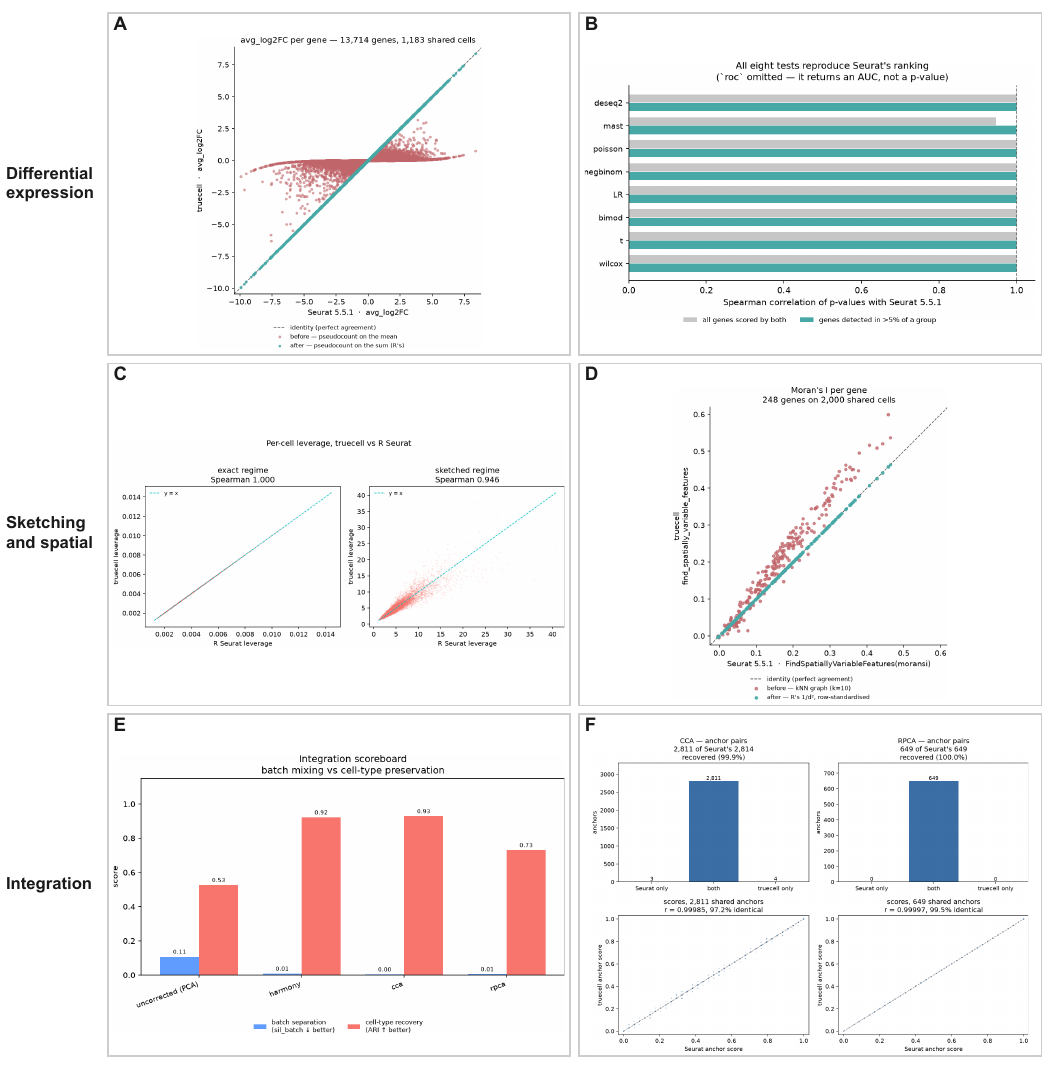}
	\end{center}
	\caption{\textbf{Measured agreement with R Seurat across analysis domains.}
		\textbf{(A)}~\texttt{avg\_log2FC} per gene, before and after matching Seurat's
		pseudocount convention. \textbf{(B)}~Rank correlation of $p$-values with
		Seurat across differential expression tests. \textbf{(C)}~Per-cell leverage
		scores in both sketching regimes. \textbf{(D)}~Moran's~I per gene on a Xenium
		slide, before and after matching Seurat's inverse-square spatial weights.
		\textbf{(E)}~Batch-integration scoreboard across correction methods.
		\textbf{(F)}~Integration anchors: how many of Seurat's anchor pairs
		Truecell finds, and how closely the shared anchors' scores agree. Each panel
		is the figure Truecell's tutorials draw at release 2.0.0, as committed to its
		repository and unaltered. The red series in (A) and (D) recomputes what an
		earlier version calculated before a fix; it is not a 2.0.0 result.}
	\label{fig:agreement}
\end{figure}

\begin{figure}[tbp]
	\begin{center}
		\includegraphics[width=\linewidth,height=0.8\textheight,keepaspectratio]{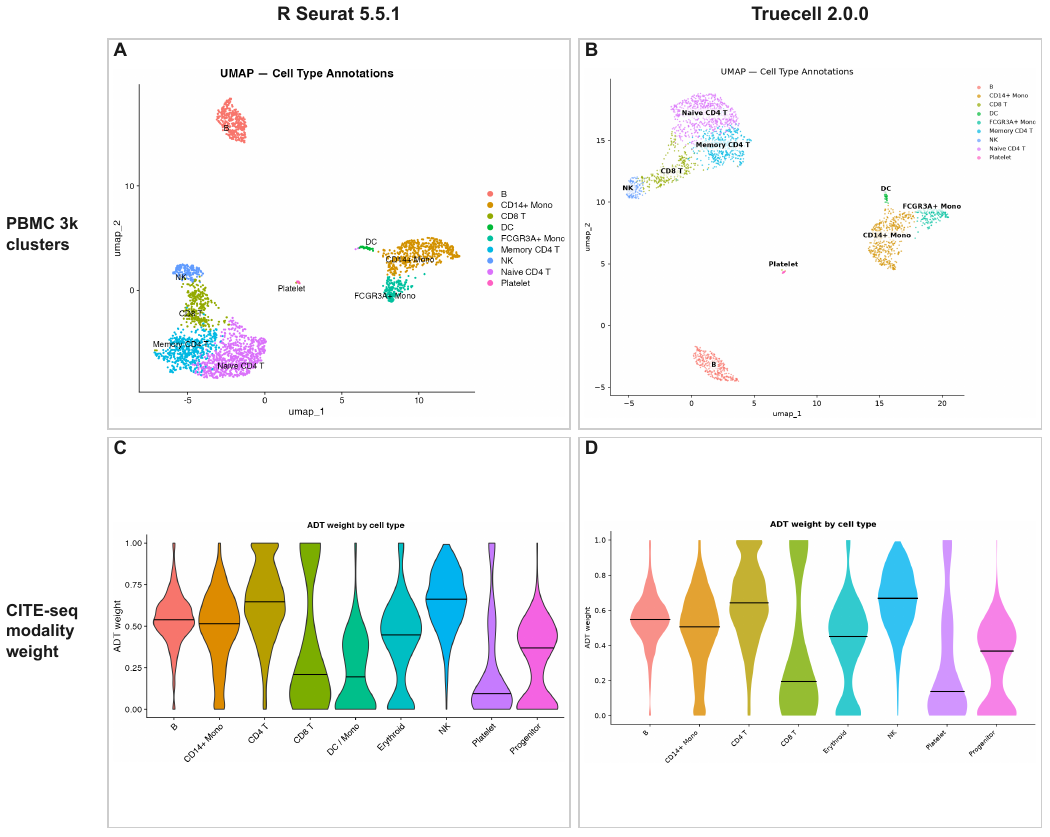}
	\end{center}
	\caption{\textbf{Representative analysis output, as each tool draws it.}
		R Seurat's figures (A,~C) beside Truecell's (B,~D), as the tutorial scripts
		draw them at commit \texttt{5dfdf29}, which changed only the drawing after
		release 2.0.0. \textbf{(A},~\textbf{B)}~The PBMC 3k guided clustering,
		labelled by cell type; the two UMAP embeddings come from different
		implementations and are not compared. \textbf{(C},~\textbf{D)}~The CITE-seq
		weighted-nearest-neighbour modality weights, by protein-defined cell type
		in one order and colour scheme; the per-cell weights correlate at Pearson
		0.984. Truecell folds Seurat's 69-cell cluster containing dendritic cells and monocytes
		  into CD14$^{+}$ monocytes on the Mac, so (D) has one group fewer;
		on Linux the tools find the same clusters.}
	\label{fig:output}
\end{figure}

Deterministic quantities agree to floating-point precision, and randomized
ones within their stated tolerance (Figure~\ref{fig:agreement}): fold changes (A),
the rank order of $p$-values in every differential expression test, with
\texttt{MAST} the exception at Spearman 0.946 (B), exact-regime leverage scores (C), Moran's~I
(D), batch mixing after integration (E) and integration anchors (F). The per-cell
protein weights of weighted-nearest-neighbour analysis correlate at Pearson 0.984
(Figure~\ref{fig:output}C,~D).

Principal components on PBMC 3k correlate with R's at a mean $|r|$ of
0.9986, and the residual comes from the two variable features the tools select
differently (Supplementary Table S6). The guided clustering reaches an adjusted
Rand index of 0.928 with 9 clusters in each tool; its B-cell, dendritic-cell and
platelet clusters are identical, and 74 of the 76 cells placed differently lie
among the T-cell and NK clusters (Supplementary Table S7).

\subsection{Rank, threshold and parameter range: the quantities an analyst
		reads off the output}
\label{sec:readable}

A Jaccard index counts the first marker and the five-hundredth as the same
event, and a worst-case bound on fold change says nothing about the order the
genes come out in. Neither is what somebody annotating clusters actually looks
at, so four comparisons were added to measure what they do look at
(Materials and Methods). All four are Truecell against R Seurat, on the paired
evaluations rather than the three-arm benchmark.

\begin{figure}[tbp]
	\begin{center}
		\includegraphics[width=\linewidth,height=0.8\textheight,keepaspectratio]{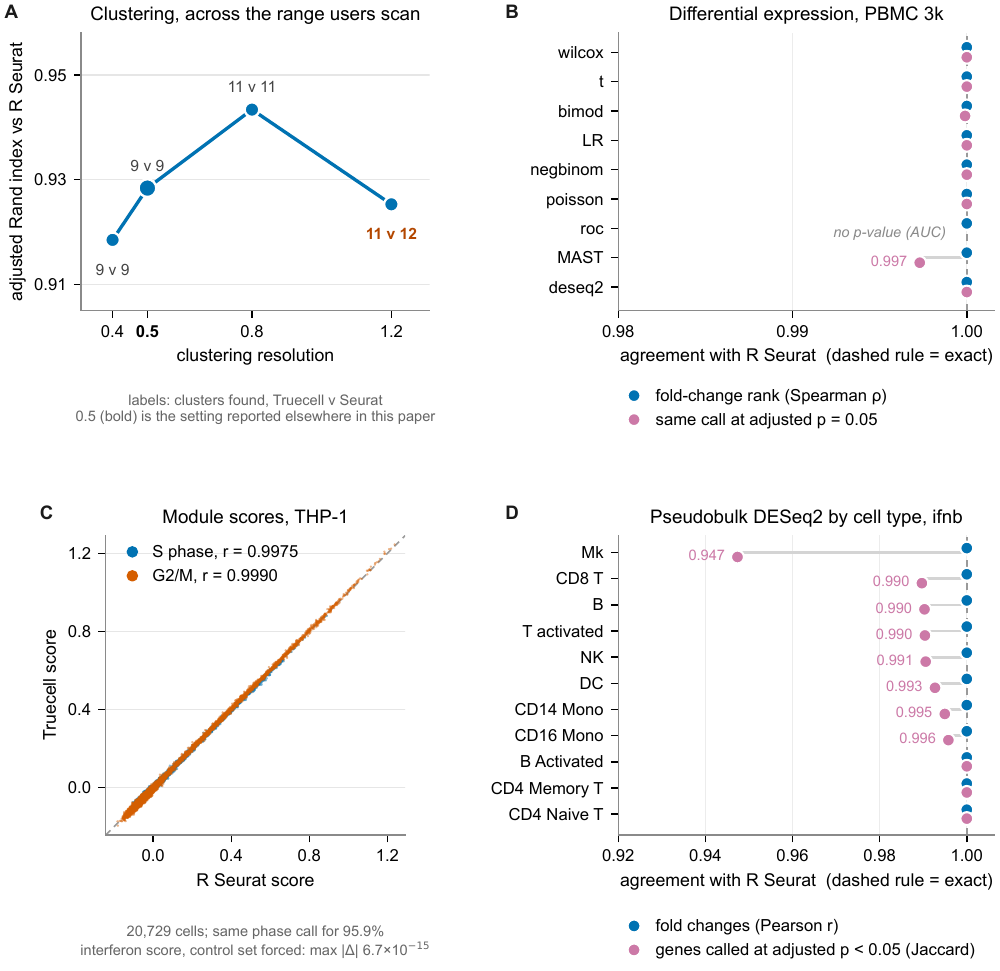}
	\end{center}
	\caption{\textbf{Agreement across the resolution range, and in the quantities
		read off a marker table, module scores and pseudobulk gene lists.}
		\textbf{(A)}~Adjusted Rand index against R Seurat on PBMC 3k at four
		clustering resolutions, both arms driven through the same vector idiom.
		Labels give the number of clusters each tool found; the counts differ
		only at 1.2. \textbf{(B)}~For each differential expression test, the rank
		correlation of \texttt{avg\_log2FC} and the fraction of genes assigned the
		same significance call at adjusted $p = 0.05$ (Table~\ref{tab:readable}).
		\texttt{roc} returns an AUC and has no threshold column. \textbf{(C)}~Cell-cycle
		S and G2/M scores per cell on THP-1 at default settings. \textbf{(D)}~Pseudobulk
		DESeq2 on ifnb, stimulated against control, by cell type: the correlation of
		fold changes and the Jaccard index of genes called at adjusted $p < 0.05$.
		The axes in (B) and (D) start above zero, so the marks are positions,
		not bar lengths.}
	\label{fig:readoff}
\end{figure}

\emph{Fold-change order is exact, and so is nearly every significance call.}
Table~\ref{tab:readable} scores the two columns an analyst uses for sorting and thresholding,
across all nine differential expression tests (Figure~\ref{fig:readoff}B).
Fold-change rank is preserved exactly for every one of the nine tests,
at Spearman and Kendall 1.000000, with the top 50 genes by
$|\log_2\mathrm{FC}|$ identical in each. Adjusted $p$-values agree to six
significant figures less often, for a reason that is not a defect: the Bonferroni
correction itself is identical, both tools computing $\min(p \times 13{,}714,\ 1)$
and agreeing on it bit for bit, but the raw $p$-values feeding it differ in their
last digits, so the product rarely lands on the same double. What survives that is
the part that decides an annotation. Every gene falls on the same side of
adjusted $p = 0.05$ in six of the nine tests, \texttt{deseq2} included;
\texttt{bimod} differs on two genes and \texttt{MAST} on 37.

\emph{Agreement holds across the resolution range.} Choosing a resolution means
running several and comparing them, so a fidelity claim pinned to one setting
says less than it appears to. Scored against R at four resolutions on PBMC 3k, the
adjusted Rand index runs 0.9185 at 0.4, 0.9284 at 0.5, 0.9434 at 0.8 and 0.9253
at 1.2 (Figure~\ref{fig:readoff}A), and the two tools find the same number of
clusters at 0.4, 0.5 and 0.8 (9, 9 and 11), while at 1.2 Truecell finds 11 and Seurat 12. These
are single-seed runs; Section~\ref{sec:benchmark} repeats them with 20 seeds.

\emph{Gene-program scores are exact once the random control set is taken
out of the comparison.} The cell-cycle evaluation in Table~\ref{tab:evals} is
a module-score result: \texttt{CellCycleScoring} is
\texttt{AddModuleScore} over the Tirosh S and G2/M sets
\citep{tirosh2016}, and on 20,729 THP-1 cells the two tools agree on 95.9\% of
phase calls, with the continuous scores correlating at Pearson 0.9975 for S
and 0.9990 for G2/M (Figure~\ref{fig:readoff}C) and at 0.9995 for an interferon
response program. Correlation is the ceiling there rather than a result, because
the control genes are sampled and R's generator is not NumPy's, and a
correlation cannot separate a faithful port from one whose binning is subtly
wrong: binning error and sampling noise land in the same residual. In the one
regime where the control set is forced rather than drawn (Materials and Methods), the two agree to $6.66\times10^{-15}$ across all 20,729 cells,
against $1.8\times10^{-1}$ between two seeds of R alone at
\texttt{ctrl}~$=$~8. The binning, the control-set selection and the mean
subtraction are therefore exact, and the residual at default settings is the
generator and nothing else. Three programs scored in one call at non-default
settings agree at Pearson 0.9972, 0.9988 and 0.9986. No exact regime exists
in the general case, which is worth stating so nobody looks for one: the bins
are uneven, and no single control count both fits the smallest and exhausts the
largest.

\emph{Group summaries agree, and pinning them found a defect.}
\texttt{aggregate\_expression} had nine tests, every one of which compared
Python against a Python re-derivation of the same formula, a form of test coverage that cannot determine whether a ported convention is incorrect. Pinned against
\texttt{AggregateExpression} for the first time, the summed counts agree
exactly, and the \texttt{return.seurat} path's normalized layer agrees to a relative tolerance of $10^{-10}$, but the first run found that Truecell had been leaving raw sums where
Seurat writes $\log(1 + \text{sums}/\text{colSums} \times 10^{4})$: 14 where
Seurat has 6.98, which any downstream function reading that layer would have
taken for the normalized expression. \texttt{average\_expression} was missing
entirely and has been added; Seurat's is not a count mean but the mean of the
back-transformed data layer, which on a small test object reads 332.84 against
a count mean of 3.17, and all three layers are now pinned to a worst absolute
difference of $6.8\times10^{-13}$ on values of the order of 300. This is
aggregation fidelity, not a pseudobulk differential expression
comparison; that comparison is in Section~\ref{sec:downstream}.

\begin{table}[tbp]
	\footnotesize
	\caption{\textbf{The two columns a marker table is read by.} Truecell
		2.0.0 against R Seurat 5.5.1 on PBMC 3k, all nine
		\texttt{find\_markers} tests, run on one shared cluster assignment so that no
		clustering difference can present itself as a differential expression
		difference. Scoring is over the 13,714 genes both arms return
		(11,387 for \texttt{negbinom} and \texttt{poisson}). ``Same call''
		is the fraction of genes on which the two tools agree about significance at
		adjusted $p = 0.05$, and the last column is the count of genes on which
		they do not. \texttt{roc} returns an AUC rather than a $p$-value, so its
		threshold columns are empty.}
	\label{tab:readable}
	\begin{center}
		\begin{tabular}{@{}lcccccc@{}}
			\toprule
			& \multicolumn{3}{c}{Fold change} & \multicolumn{3}{c}{Adjusted $p$} \\
			\cmidrule(lr){2-4}\cmidrule(lr){5-7}
			Test & Spearman & Kendall & Top 50 by $|\log_2\mathrm{FC}|$ & To 6 s.f. & Same call & Genes differing \\
			\midrule
			\texttt{wilcox}       & 1.000000 & 1.000000 & 50/50 & 0.9932 & 1.0000 & 0 \\
			\texttt{t}            & 1.000000 & 1.000000 & 50/50 & 1.0000 & 1.0000 & 0 \\
			\texttt{bimod}        & 1.000000 & 1.000000 & 50/50 & 0.9996 & 0.9999 & 2 \\
			\texttt{LR}           & 1.000000 & 1.000000 & 50/50 & 0.9909 & 1.0000 & 0 \\
			\texttt{negbinom}     & 1.000000 & 1.000000 & 50/50 & 0.9809 & 1.0000 & 0 \\
			\texttt{poisson} & 1.000000 & 1.000000 & 50/50 & 0.8538 & 1.0000 & 0 \\
			\texttt{roc}          & 1.000000 & 1.000000 & 50/50 & --- & --- & --- \\
			\texttt{MAST}         & 1.000000 & 1.000000 & 50/50 & 0.8556 & 0.9973 & 37 \\
			\texttt{deseq2}       & 1.000000 & 1.000000 & 50/50 & 0.9415 & 1.0000 & 0 \\
			\bottomrule
		\end{tabular}
	\end{center}
\end{table}

\subsection{Downstream of feature selection, Truecell tracks Seurat more closely
	than Scanpy, and the remaining gap is feature selection}
\label{sec:benchmark}

\begin{figure}[tbp]
	\begin{center}
		\includegraphics[width=\linewidth,height=0.8\textheight,keepaspectratio]{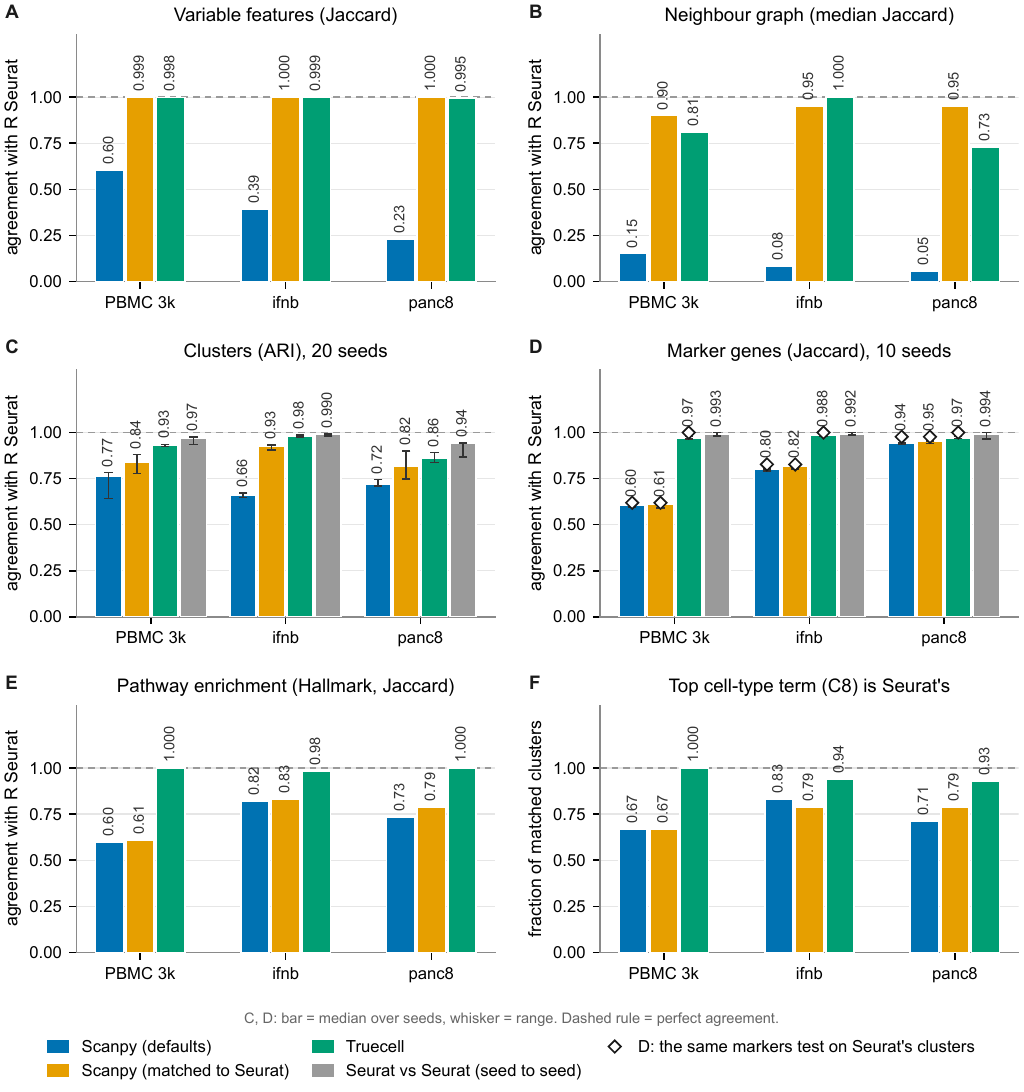}
	\end{center}
	\caption{\textbf{Three-arm benchmark against R Seurat 5.5.1 on three datasets.}
		Each panel is one metric; bars are the three comparison arms; the dashed rule
		indicates perfect agreement with the reference. Every parameter a user sets
		explicitly is identical across arms, so the differences shown are between
		implementations rather than between default arguments. Scanpy appears twice:
		at its own defaults, and configured to imitate Seurat's variable-feature
		selection and clustering. \textbf{(C)}~Median over 20 seeds, with the
		range; grey, Seurat's own seeds against its other 19. \textbf{(D)}~Markers on
		each arm's own clustering over seeds 1 to 10, with the range; diamonds, the
		same test on Seurat's clusters. \textbf{(E)}~Median Jaccard index of the
		Hallmark sets significant for matched clusters' markers.
		\textbf{(F)}~Fraction of matched clusters whose top C8 cell-type term is
		Seurat's.}
	\label{fig:benchmark}
\end{figure}

Agreement with a reference is only informative in relation to alternatives. We
therefore repeated the comparison \citet{rich2026} report, with three arms
rather than two (R Seurat, Scanpy and Truecell) on three datasets: 10x
PBMC 3k (2,638 cells after quality control), the interferon-$\beta$ stimulation
dataset \citep[ifnb, 13,999 cells;][]{kang2018}, and the panc8 compendium of
human islet cells (14,890 cells). Every parameter a user sets explicitly was
held equal across arms, so that what is measured is each implementation rather
than a difference in the default arguments. Scanpy was run twice: once at its
own defaults, and once configured to imitate Seurat as closely as its API allows,
using its \texttt{seurat\_v3} variable-feature method and Louvain clustering.
Every arm, Seurat included, was clustered with 20 seeds at four resolutions
(Materials and Methods). The results are presented in
Figure~\ref{fig:benchmark} and Table~\ref{tab:benchmark}.

\begin{table}[tbp]
	\footnotesize
	\caption{\textbf{Three-arm benchmark, all metrics.} Agreement with R Seurat
		5.5.1; higher values are closer to the reference. Neighbour graph:
		scored against an exact search on Seurat's embedding. Clusters: at resolution
		0.5, the median over an arm's 20 seeds of its mean adjusted Rand index
		against Seurat's 20 seeds, with the range; for Seurat, each seed against the other 19. Markers: the union of positive markers from each arm's canonical
		run and, under ``common'', from each arm's test on Seurat's most
		representative partition, where Scanpy's two arms are one computation. The
		lower block names the arm closest to Seurat, with its margin; margins below
		0.005, including the panc8 variable-feature margin of 0.00499, are ties.}
	\label{tab:benchmark}
	\begin{center}
		\begin{tabular}{@{}llccccc@{}}
			\toprule
			& & Variable & Neighbour & Clusters, median & \multicolumn{2}{c}{Markers} \\
			\cmidrule(lr){6-7}
			Dataset & Arm & features & graph & [range] over seeds & canonical & common \\
			\midrule
			PBMC 3k & Scanpy (defaults) & 0.603 & 0.152 & 0.765 [0.642, 0.783] & 0.598 & 0.619 \\
			 & Scanpy (matched) & 0.999 & 0.900 & 0.841 [0.778, 0.881] & 0.608 & 0.619 \\
			 & Truecell & 0.998 & 0.810 & 0.932 [0.924, 0.936] & 0.970 & 1.000 \\
			 & Seurat, seed to seed & & & 0.971 [0.935, 0.976] & & \\
			\addlinespace
			ifnb & Scanpy (defaults) & 0.393 & 0.083 & 0.662 [0.648, 0.672] & 0.800 & 0.827 \\
			 & Scanpy (matched) & 1.000 & 0.950 & 0.928 [0.904, 0.933] & 0.803 & 0.827 \\
			 & Truecell & 0.999 & 1.000 & 0.981 [0.975, 0.986] & 0.985 & 1.000 \\
			 & Seurat, seed to seed & & & 0.990 [0.980, 0.993] & & \\
			\addlinespace
			panc8 & Scanpy (defaults) & 0.232 & 0.054 & 0.722 [0.712, 0.745] & 0.944 & 0.977 \\
			 & Scanpy (matched) & 1.000 & 0.950 & 0.816 [0.749, 0.899] & 0.971 & 0.977 \\
			 & Truecell & 0.995 & 0.727 & 0.859 [0.838, 0.891] & 0.989 & 1.000 \\
			 & Seurat, seed to seed & & & 0.942 [0.867, 0.943] & & \\
			\midrule
			\multicolumn{7}{@{}l}{\emph{Closest arm (margin)}} \\
			PBMC 3k & & tie & matched (0.090) & Truecell (0.091) & Truecell (0.362) & \\
			ifnb & & tie & Truecell (0.050) & Truecell (0.052) & Truecell (0.182) & \\
			panc8 & & tie & matched (0.223) & Truecell (0.043) & Truecell (0.018) & \\
			\bottomrule
		\end{tabular}
	\end{center}
\end{table}

Configured to match Seurat, Scanpy repairs its variable features and most of
its neighbour graph (Figure~\ref{fig:benchmark}A,~B). Given Seurat's variable
features, Truecell's graph matches the exact reference on every dataset, so its
graph gap is feature selection alone (Supplementary Table S6).

Truecell is closer to Seurat's clustering than the Seurat-configured Scanpy
in all 12 combinations of dataset and resolution, significantly so in each (Holm-adjusted
$p < 0.001$; Figure~\ref{fig:benchmark}C, Supplementary Figure~S1 and Table~S3).
At resolution 0.5 the Hodges--Lehmann difference is 0.092 on PBMC 3k, 0.055 on ifnb
and 0.048 on panc8. Seurat does not reproduce itself exactly either: its
seeds agree with one another at a median of 0.971, 0.990 and 0.942, and Truecell
stays below that ceiling in all 12 settings. In an exploratory analysis,
when given Seurat's variable features, Truecell agrees with Seurat as closely as
Seurat's own seeds do (Supplementary Note~S2): the remaining gap comes from the
few variable features that differ.

Truecell's marker genes are also closer to Seurat's than either Scanpy
arm's on every dataset, whatever the seed, and on Seurat's own clusters its test
returns Seurat's markers exactly (Figure~\ref{fig:benchmark}D; Supplementary
Table S5). Over the four metrics and three datasets, Truecell is closest to
Seurat in 7 of 12 comparisons, the Seurat-configured Scanpy in two, and three are
ties (Table~\ref{tab:benchmark}).

Feature selection is therefore where Truecell and Seurat still differ, and
the difference is deterministic. Truecell's variable features differ from
Seurat's by two genes of 2,000 on PBMC 3k, one on ifnb and five on panc8.
Not every variable feature carries the same weight, so which genes differ is a
separate question from how many, and it is answerable here by inspection.
On PBMC 3k, \emph{GINS2} and \emph{WBP5} drop out and \emph{ACTG1} and
\emph{CYBA} come in, none of them a lineage or activation marker. On ifnb
exactly one gene differs across 13,999 cells, and it runs in Truecell's favour:
the gene gained is \emph{CD3G}, a component of the T-cell receptor complex, and
the one dropped is \emph{CCNB1IP1}, a cell-cycle-associated E3 ligase.
The panc8 dataset is the adverse case: the genes lost include the islet
markers \emph{SERPINA1} and \emph{CRYBA2}, which is why this is reported as a
limitation rather than a curiosity. Seurat obtains its variance-stabilizing fit
from R's \texttt{loess} with its default kd-tree interpolated surface; Truecell
computes the same fit exactly. The two agree closely over most of the expression
range and diverge where the data are sparsest, which is enough to move a handful
of highly expressed genes across the 2,000-feature cutoff. Feature selection
has no random seed in either tool, so this is not a seeding artefact, and we
report it as a limitation rather than resolving it: matching Seurat here would
mean reproducing an approximation rather than computing the fit, and doing so
partially would be worse than either alternative.

\subsection{Pathway readouts and pseudobulk gene lists match Seurat's}
\label{sec:downstream}

Marker genes are usually read through enrichment. For every dataset and
gene-set collection, Truecell's clusters point to the same pathways as Seurat's
more often than either Scanpy arm's (Figure~\ref{fig:benchmark}E; Supplementary
Table S8). With the Hallmark sets the median Jaccard index is 1.000, 0.984 and
1.000, against 0.611, 0.831 and 0.790 for the Seurat-configured Scanpy, and all
18 paired comparisons with the Scanpy arms are significant (Holm-adjusted
$p \leq 0.031$). The top C8 cell-type term is
Seurat's for 37 of 39 matched Truecell clusters, against 28 of 37 for the
Seurat-configured Scanpy (Figure~\ref{fig:benchmark}F).

Pseudobulk DESeq2 on ifnb reproduces Seurat's result
(Figure~\ref{fig:readoff}D; Supplementary Table S9). The two implementations produce identical counts across all 160 pseudobulk samples, fold changes agree to within $10^{-14}$, and the genes
called at adjusted $p < 0.05$ agree with a Jaccard index of 0.947 to 1.000 across
the 11 cell types (median 0.993). Both tools find the Hallmark interferon
response sets significant in all 11.

\subsection{Spatial transcriptomics}
\label{sec:spatial}

Truecell reads Xenium, Visium, CosMx and MERSCOPE output; Supplementary
Table S14 gives what has been verified against R for each. On the public Xenium human
breast cancer section, the two tools load the same 166,363 cells and 313 genes
with identical counts and centroids, principal components correlate at a mean
$|r|$ of 0.999999996, and Moran's~I agrees to $2.2\times10^{-14}$; five seeds of
each tool agree at a median adjusted Rand index of 0.934, against 0.955 between
Seurat's own seeds (Supplementary Table S11). The comparison exposed a defect in Truecell's \texttt{subset},
which has since been fixed: given cells in a different order, it misaligned the object's slots
(Supplementary Table S12). Truecell objects convert to AnnData with the
coordinates in \texttt{obsm["spatial"]}, and a round trip through a SpatialData
store is exact.

\subsection{Reproducing Seurat costs time on the standard workflow and
	saves it on most individual operations}
\label{sec:performance}

\begin{figure}[tbp]
	\begin{center}
		\includegraphics[width=\linewidth,height=0.8\textheight,keepaspectratio]{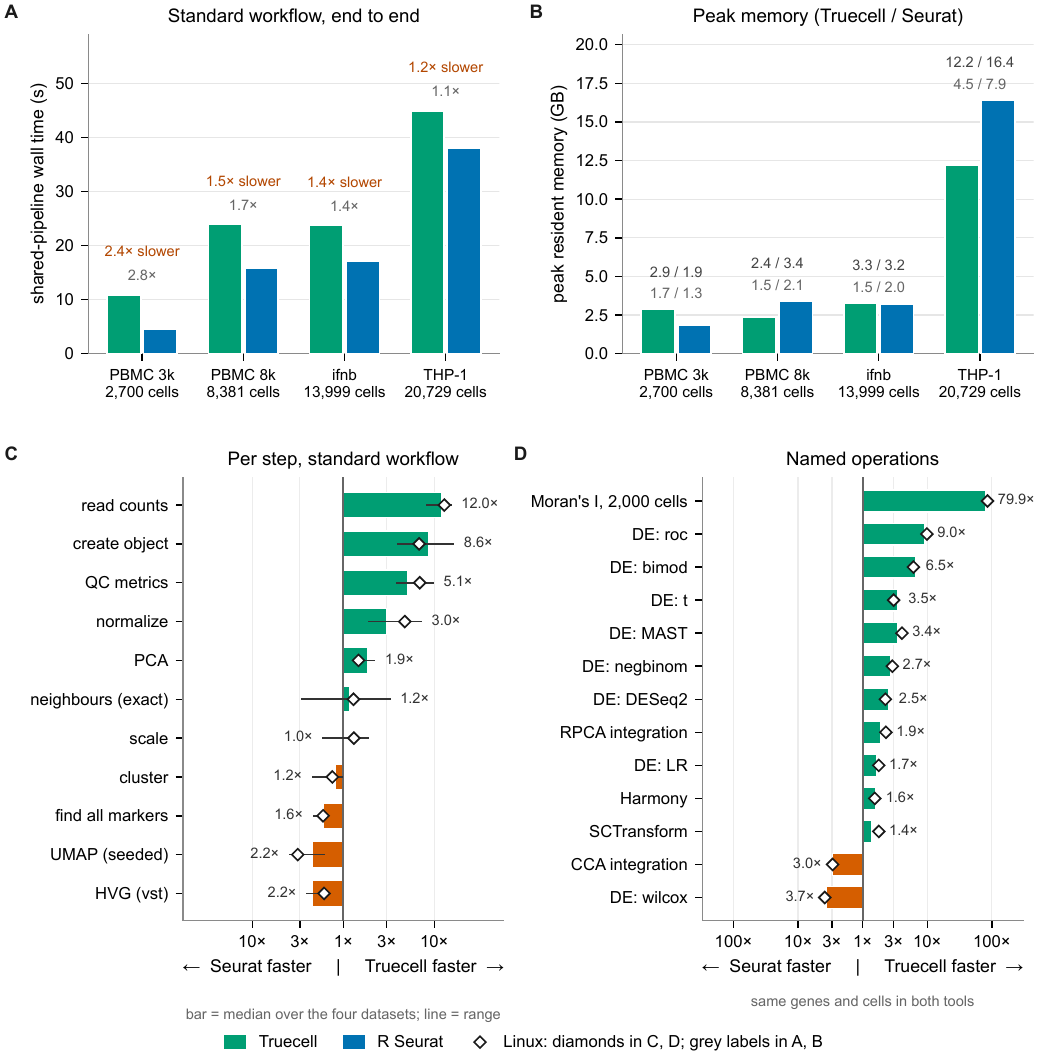}
	\end{center}
	\caption{\textbf{Runtime and peak memory against R Seurat 5.5.1, on two
		machines.} Both arms run line-for-line counterpart scripts on the same inputs,
		with both language runtimes on an optimized BLAS library, and record an anchor at
		every step so that only like-for-like work is compared. Bars show the results from the Apple
		M5 Pro; the Linux x86-64 server is shown by the grey labels in (A) and (B) and
		the open diamonds in (C) and (D). \textbf{(A)}~Wall time for the standard
		clustering workflow, excluding interpreter start-up and \texttt{library()}.
		\textbf{(B)}~Peak resident memory for the same runs, in binary
		gigabytes (1~GB $=$ 1,024~MB). \textbf{(C)}~Per-operation ratio across those
		four datasets: the bar is the median, the line shows the full range, and the axis is logarithmic so that a loss and a win of equal magnitude are equally legible.
		\textbf{(D)}~The same ratio for named operations: SCTransform and eight
		differential expression tests on PBMC 3k, integration on ifnb, and
		Moran's~I on 2,000 cells of the Xenium mouse brain section.}
	\label{fig:performance}
\end{figure}

On the standard workflow, from reading counts to marker detection, Truecell
took 1.2 to 2.4 times as long as R Seurat on the Mac and 1.1 to 2.8 times as
long on Linux, the ratio falling as datasets grow, and both tools found the same
number of clusters and marker rows on every dataset
(Figure~\ref{fig:performance}A; Supplementary Tables S10 and S10c). Truecell's
peak memory was lower than Seurat's on PBMC 8k and THP-1 and higher on PBMC 3k on
both machines (Figure~\ref{fig:performance}B).

Individual operations mostly favour Truecell: it was faster in 44 of 68
like-for-like step comparisons on the Mac and 45 on Linux, including seven of the
eight differential expression tests, and Moran's~I on 2,000 Xenium cells ran 80
times faster on the Mac and 86 times faster on Linux
(Figure~\ref{fig:performance}C,~D).
Two steps explain the end-to-end loss: seeded UMAP, which \texttt{umap-learn}
runs on a single thread, and the Wilcoxon test inside marker detection (3.7 times
slower on the Mac, 3.9 on Linux), which Seurat runs through the \texttt{Presto} C++ implementation
\citep{korsunsky2019presto}. Both are differences between libraries, not between
languages. Seurat's SCTransform ran without its optional
\texttt{glmGamPoi} accelerator.

\subsection{Comparison against the reference exposed defects on both
	sides}
\label{sec:defects}

Building evaluations was not only a matter of confirming agreement. Eleven of the eighteen evaluations surfaced defects in Truecell that its own unit tests had not: a
normalization applied along the wrong margin, a differential expression
statistic that used a moment-based dispersion where Seurat uses a
maximum-likelihood one, an out-of-core code path that densified the matrix it
was supposed to stream, and others. Each was found because a number disagreed
with R at a tolerance tighter than any internal test, and each was fixed
before a release.

The analyses added for this paper found seven more in Truecell 1.2.0,
in subsetting, marker defaults, the DESeq2 path, clustering restarts, AnnData
coordinates and gene renaming, all fixed in 2.0.0 (Supplementary Table S12). A disagreement with an identified cause on Truecell's side is a
defect to fix, not a caveat to disclose, because agreement with Seurat is the
port's purpose.

The same tolerance also applies in the other direction. The Visium evaluation
found that Seurat's \texttt{VisiumV2} container stores the spot diameter in the
field its own accessor documents as a radius, and that calling
\texttt{Radius()} on that object returns \texttt{NULL}; the slide's fixed
100~\textmu m spot pitch settles which of the two quantities was stored:
read as a radius, the stored value would describe spots 130.6~\textmu m across on
a 100~\textmu m grid. Truecell stores the radius, because copying Seurat here would make it agree
with Seurat and disagree with the data, and a test pins the divergence. Both
behaviours are present in Seurat 5.5.1. The \texttt{NULL} was a defect, and
another contributor has since fixed it in Seurat's development version
(pull request \href{https://github.com/satijalab/seurat/pull/10454}{satijalab/seurat\#10454}). The stored diameter is
what Seurat's own tests expect, so we did not report it as a defect to the Seurat developers, and Truecell's
documentation states the difference. A comparison of this kind is a test of both
implementations, and finding a discrepancy does not by itself say which
side is wrong; that has to be established case by case, as it was here.

\subsection{Skill capsules tell AI coding agents how to use Truecell}
\label{sec:skills}

An AI coding agent, such as Claude Code, is a program that uses a language
model to write and run code at a user's request. Before writing code for a
library, an agent can load a \emph{skill}: a short Markdown file whose header
says when the file applies and whose body explains how to use the library.
Truecell 2.0.0 includes ten such files, which we call skill capsules. One covers installation and the standard pipeline and
points to the other nine, each written for one task, such as differential
expression, integration, spatial data or porting Seurat code. Each capsule
lists the functions its task needs, with their arguments, defaults and return
values. Being plain Markdown, the capsules also work with other language
models.

The capsule for porting Seurat code is an example. An agent reads its
header to decide whether to load the file; the body includes a map from
Seurat's functions to Truecell's. The header and five of the map's 38 rows
are:

\begin{center}
	\footnotesize
	\begin{tabular}{@{}>{\raggedright\arraybackslash}p{0.15\linewidth}>{\raggedright\arraybackslash\ttfamily}p{0.39\linewidth}>{\raggedright\arraybackslash\ttfamily}p{0.39\linewidth}@{}}
		\toprule
		\multicolumn{3}{@{}l}{\textbf{Header} of \texttt{skills/truecell-from-seurat/SKILL.md}} \\
		\midrule
		\texttt{name:} & \multicolumn{2}{>{\raggedright\arraybackslash\ttfamily}p{0.80\linewidth}@{}}{truecell-from-seurat} \\
		\texttt{description:} & \multicolumn{2}{>{\raggedright\arraybackslash\ttfamily}p{0.80\linewidth}@{}}{Use when porting R Seurat code to Python truecell, translating a Seurat function or argument name, or comparing the two tools\textquotesingle{} numeric output. [\ldots]} \\
		\midrule
		\multicolumn{3}{@{}l}{\textbf{Function map} (5 of 38 rows)} \\
		\midrule
		\textbf{Task} & \textbf{\textrm{R (Seurat)}} & \textbf{\textrm{Python (truecell)}} \\
		Normalize & NormalizeData(obj, normalization.method, scale.factor) & normalize\_data(obj, normalization\_method=, scale\_factor=) \\
		HVGs & FindVariableFeatures(obj, selection.method, nfeatures) & find\_variable\_features(obj, selection\_method=, nfeatures=) \\
		Neighbours & FindNeighbors(obj, dims=1:10) & find\_neighbors(obj, dims=range(10)) \\
		Cluster & FindClusters(obj, resolution) & find\_clusters(obj, resolution=, algorithm=) \\
		All markers & FindAllMarkers(obj, only.pos, logfc.threshold) & find\_all\_markers(obj, only\_pos=, logfc\_threshold=) \\
		\bottomrule
	\end{tabular}
\end{center}

With this map, an agent asked to run a Seurat script in Python can
translate the script one call at a time. The capsule ends with a worked
example, the PBMC 3k tutorial in R and in Python, and both versions keep the
same 2,638 cells after quality control (Table~\ref{tab:evals}).

Truecell's automated tests, run on every proposed change, check the
capsules against the code. They fail if a public function is not mentioned in
the skill files, or if any of the 74 function signatures in the capsules' API
reference differs from the code in its parameters, their order or their
defaults.

An agent that writes analysis code is only worth using if the code produces a
result that a competent analyst would have produced. The capsules help an
agent call the right functions with the right arguments, and
Sections~\ref{sec:evals} to~\ref{sec:benchmark} show that those functions
reproduce Seurat, a reference whose behaviour is documented and widely
understood.

\section{Discussion}

Truecell is not proposed as a replacement for Scanpy, and the two are not in
competition for the same user base. Scanpy is a mature, well-engineered
platform at the centre of a large ecosystem, and its GPU-accelerated lineage is
substantially faster than Seurat \citep{billato2025} or Truecell. The runtime
comparison we do report is narrower on every dimension: Truecell against
Seurat, on CPUs, at 2,700 to 20,729 cells. Nor is Seurat's answer the
standard against which biological correctness should be judged:
Bioconductor-based pipelines outscore both Seurat and Scanpy on clustering
accuracy in a recent benchmark \citep{billato2025}. The claim here is
narrower and, we think, more useful: for a group that already has a Seurat
analysis and needs to run it in Python, Truecell reproduces that analysis more
closely than the alternatives we tested and does so without requiring it
to be rewritten.

Format conversion does not solve the same problem. Moving a Seurat object into
AnnData and continuing in Scanpy transfers the data, but not the analysis:
every call has to be rewritten and its defaults re-derived, which is precisely
the step that \citet{rich2026} demonstrate alters the result. Defaults do not
travel with the data. Software already exists to move the data itself between
the two ecosystems \citep{feng2022}, which addresses a problem complementary to the one considered here.

The Scanpy-matched arm is corroborating evidence for this view rather than just
a comparison point. Scanpy's \texttt{seurat\_v3} variable-feature method is
itself a port of one component of Seurat, its \texttt{vst} feature selection,
and it reproduces Seurat's feature set almost exactly on all three datasets
tested here, which is the closest any comparator arm gets to Seurat on any
metric. That a second, independently written port of a single Seurat algorithm
also achieves near-exact fidelity is evidence for porting as a strategy, not only
for Truecell as an implementation of it.

The difference between re-deriving and porting reaches the pathways an
analyst reads: the median Hallmark Jaccard index was 0.61 to 0.83 for the
Seurat-configured Scanpy and 0.98 to 1.00 for Truecell.

The performance result defines an important trade-off: a group making the transition gains fidelity but incurs a wall-time cost on the standard workflow, by a factor that falls from 2.4 at 2,700 cells
to 1.2 at 20,729 on the Mac, and from 2.8 to 1.1 on Linux, while most
individual operations run faster in Truecell.

\subsection{Limitations}
\label{sec:limitations}

The results are limited in six ways.
\begin{enumerate}
	\item \emph{Fidelity is not correctness.} Agreement is with R Seurat as a
	reference implementation.
	\item \emph{One reference version.} Seurat's output changes between versions
	\citep{rich2026}; the reference here is Seurat 5.5.1.
	\item \emph{Feature selection.} One to five of the 2,000 variable features
	differ, traced to R's interpolated \texttt{loess} surface; this accounts for the
	remaining neighbour-graph and clustering gaps.
	\item \emph{Benchmark scope.} Three datasets, four resolutions, 10 principal
	components and $k = 20$.
	\item \emph{Downstream scope.} Over-representation enrichment and one
	pseudobulk design, on one dataset.
	\item \emph{Platforms.} One macOS machine and one 8-core virtual Linux server,
	against R Seurat only; the CosMx and MERSCOPE loaders are checked against R on
	test files only.
\end{enumerate}

The broader observation is that comparison at this tolerance is informative
about both implementations. Several of the discrepancies we chased turned out
not to be defects in the port but properties of the reference that its own
users have no particular reason to know: a default \texttt{loess} surface that
interpolates, a container field whose name and contents disagree. Reproducing a
widely used tool closely enough to find such things seems a reasonable
definition of what a port should achieve.

\section*{Conflict of Interest Statement}
The authors declare that the research was conducted in the absence of any
commercial or financial relationships that could be construed as a potential
conflict of interest.

\section*{Generative AI Statement}
Generative AI was used during manuscript preparation and software development. Claude (Anthropic) assisted with drafting and revising portions of the text, including an earlier version of the Figure 1 caption, and with generating and reviewing code. All AI-assisted text was reviewed and edited by the authors. All analyses were executed using the paper's analysis scripts, and all reported results were independently inspected and verified by the authors. No AI tool is listed as an author.

\section*{Author Contributions}
N.I.: conceptualization, writing, review and editing. S.A.:
conceptualization, writing, review and editing. M.G.: review and editing.
S.C.G.: review and editing.

\section*{Funding}
This work received no external funding. 

\section*{Acknowledgments}

We thank the developers of Seurat, Scanpy, NumPy, SciPy, and scikit-learn.
Truecell is an independent project and is not affiliated with, endorsed by, or
maintained by the Satija Lab or 10x Genomics.

\section*{Supplementary Material}

Supplementary Tables S1 to S14, Supplementary Figure S1, Supplementary
Notes S1 to S3 and Supplementary Data Sheet 1 are provided in the Supplementary Material.

\section*{Data Availability Statement}

All datasets analysed in this study are publicly available and are listed in
Supplementary Table S13. Truecell itself, the paired R and Python scripts for the eighteen evaluations and the performance benchmark suite are available at
\url{https://github.com/GenomicAI/truecell} under the MIT licence, with
documentation at \url{https://genomicai.github.io/truecell/}.

\bibliographystyle{Frontiers-Harvard}
\bibliography{bibliography}

\end{document}